\documentclass[sigconf,nonacm]{acmart}
\usepackage{float}
\usepackage{subfig}
\usepackage{tabularx}\usepackage{hyperref}
\usepackage[linesnumbered,ruled,vlined]{algorithm2e}

\SetAlgorithmName{Algorithm}{Algorithm}{List of Algorithms}

\newcommand\vldbdoi{XX.XX/XXX.XX}
\newcommand\vldbpages{XXX-XXX}
\newcommand\vldbvolume{14}
\newcommand\vldbissue{1}
\newcommand\vldbyear{2020}
\newcommand\vldbauthors{\authors}
\newcommand\vldbtitle{\shorttitle} 
\newcommand\vldbavailabilityurl{https://github.com/umr-dbs/cMVBT}
\newcommand\vldbpagestyle{plain} 

\begin{document}
\title{Multiversion Concurrency Control for \\Multiversion B-Trees}

%%
%% The "author" command and its associated commands are used to define the authors and their affiliations.
\author{Amir Tonta}
\affiliation{%
  \institution{University of Marburg}
  \city{Marburg}
  % \state{Hesse}
  \country{Germany}
}
\email{amir.tonta@mathematik.uni-marburg.de}
\orcid{0009-0007-4561-9850}

\author{Bernhard Seeger}
\affiliation{%
  \institution{University of Marburg}
  \city{Marburg}
  % \state{Hesse}
  \country{Germany}
}
\email{seeger@informatik.uni-marburg.de} % This should match your CMT Account specifically.
\orcid{0000-0002-9362-153X}

\author{Eljas Soisalon-Soininen}
\affiliation{%
  \institution{Aalto University}
  \city{Espoo}
  \country{Finland}
}
\email{eljas.soisalon-soininen@aalto.fi}
\orcid{0000-0001-6437-2127}

\renewcommand{\subsectionautorefname}{Section}
\renewcommand{\subsubsectionautorefname}{Section}

\begin{abstract}
Multiversion concurrency control (MVCC) enables scans to read data from a committed snapshot (version), reducing conflicts with write operations compared to traditional concurrency approaches. Currently, versioned records are often managed in a B$^+$-tree using version chains. However, version chains introduce overhead during scans and can still lead to conflicts between scans and writers. The multiversion B-tree (MVBT) was designed for optimal range scan performance on arbitrary versions, but has been considered impractical due to its structural complexity and, until recently, the lack of effective concurrency control. In this paper, we present the concurrent MVBT (cMVBT), a redesign of the MVBT featuring a novel concurrency control protocol that uses optimistic latches for write operations and requires no latches for range scans, while preserving all the optimality guarantees of the original MVBT. Additionally, cMVBT supports continuous garbage collection without activity spikes, seamlessly integrating free-space management. Experiments with mixed workloads derived from a standard benchmark show that the cMVBT achieves low overhead, high write throughput, and excellent range scan performance, outperforming state-of-the-art methods based on version chains.

% Multiversion concurrency control (MVCC) enables read operations to access consistent snapshots without blocking writers. Common data structures, such as B$^+$-tree extensions with version support, often rely on version chains. However, these chains introduce overhead during scans and garbage collection, and can still lead to conflicts between readers and writers. The multiversion B-tree (MVBT) was designed specifically for multi-versioning but has been considered impractical due to its structural complexity and, until now, the absence of effective concurrency control.  In this paper, we present new concurrency control protocols for MVBT that use optimistic latches for write operations while achieving optimal range-scan performance without any latching. Additionally, our concurrent MVBT (cMVBT) enables continuous garbage collection without activity spikes and incorporates free-space management. Experiments on a standard benchmark show that the cMVBT achieves low overhead, high write throughput, and efficient range scan processing, outperforming state-of-the-art methods based on version chains.
\end{abstract}

\maketitle

\pagestyle{\vldbpagestyle}
\begingroup\small\noindent\raggedright\textbf{PVLDB Reference Format:}\\
\vldbauthors. \vldbtitle. PVLDB, \vldbvolume(\vldbissue): \vldbpages, \vldbyear.\\
\href{https://doi.org/\vldbdoi}{doi:\vldbdoi}
\endgroup
\begingroup
\renewcommand\thefootnote{}\footnote{\noindent
This work is licensed under the Creative Commons BY-NC-ND 4.0 International License. Visit \url{https://creativecommons.org/licenses/by-nc-nd/4.0/} to view a copy of this license. For any use beyond those covered by this license, obtain permission by emailing \href{mailto:info@vldb.org}{info@vldb.org}. Copyright is held by the owner/author(s). Publication rights licensed to the VLDB Endowment. \\
\raggedright Proceedings of the VLDB Endowment, Vol. \vldbvolume, No. \vldbissue\ %
ISSN 2150-8097. \\
\href{https://doi.org/\vldbdoi}{doi:\vldbdoi} \\
}\addtocounter{footnote}{-1}\endgroup

\ifdefempty{\vldbavailabilityurl}{}{
\vspace{.3cm}
\begingroup\small\noindent\raggedright\textbf{PVLDB Artifact Availability:}\\
The source code, data, and/or other artifacts have been made available at \url{github.com/umr-dbs/cMVBT} and \url{github.com/umr-dbs/BTree-MVCC-Version-Chains}.
\endgroup
}

\section{Introduction}
Efficient multiversion concurrency control (MVCC) is essential for high-performance database systems 
\cite{wu2017empirical}, particularly those such as Hyper \cite{kemper2011hyper} that support hybrid transaction/analytical processing (HTAP) workloads \cite{zhang2024htap}. By allowing queries to run on older versions (snapshots) while updates modify the most recent version, MVCC offers snapshot isolation and robust parallelism without operational conflicts. A critical component is an index structure that supports single-row write operations (insertions, deletions, and updates) and range scans concurrently across snapshots. Designing such structures remains challenging due to the difficulty of managing multiple versions efficiently under concurrent workloads.

There are two fundamentally different approaches to indexing for version management. First, the most common approach relies on version lists attached to each key in a B$^+$-tree \cite{wu2017empirical}. Each version list contains entries with timestamps and values, organized chronologically, often with the most recent entry first. Single-row write operations append new versions to the front of the list, e.g., a deletion corresponds to inserting a tombstone record. Second, the other approach employs a copy-on-write B$^+$-tree with path copying, as used for example in CouchDB \cite{DBLP:books/daglib/0024051}. A write operation triggers the insertion of a new path, while a scan query runs without acquiring latches by first determining the root of a committed version on which the scan is performed.  However, both approaches come with their own drawbacks.   

%The standard indexing approach to version management often relies on version lists attached to each key in a B$^+$-tree. Each version list contains entries with timestamps and values, organized chronologically, with the most recent entry first. Updates append new versions to the front of the list, and deletions add tombstone entries but do not physically remove keys from the B$^+$-tree. New key insertions create a list with a single entry. Range scans on a given version, typically the last committed version, first access the B$^+$-tree leaves and then traverse the version lists of each matching key. 

For the list-based approach, scan queries slow down as the lists grow, leading to a vicious circle of steadily declining performance, as outlined in \cite{bottcher2019scalable}. Although clever garbage collection and more advanced list organization, such as frugal skiplists and vWeaver \cite{kim2021rethink}, mitigate the retrieval problem, they can incur substantial overhead for range scans for mixed workloads. Second, deletions degrade performance because keys often remain in the B$^+$-tree even after they are logically deleted \cite{johnson1993performance}. Third, range scans can still conflict with concurrent insertions in highly contended pages, thereby reducing query performance. These concurrency problems with high contention might be mitigated by using lightweight latches for B$^+$-trees \cite{bottcher2020scalable, el2024lightweight}, which provide effective mechanisms for coping with high contention in conflict-prone workloads.  However, latch efficiency alone does not eliminate scalability concerns under mixed workloads. 

In CoW-based approaches, write amplification is high because each write operation triggers a copy of a B$^+$-tree path \cite{DBLP:books/daglib/0024051}. Consequently, garbage collection becomes more expensive and more demanding. In addition, single-write transactions must be performed sequentially. This problem can be alleviated by batching and performing batches in parallel \cite{DBLP:journals/pvldb/0001BLP19}. However, this can increase latency, while the effort for space management and garbage collection remains high. 

% This list-based approach causes multiple problems.  First, scan queries slow down as the lists grow, leading to a vicious circle of steadily declining performance, as outlined in \cite{bottcher2019scalable}. Although clever garbage collection and more advanced list organization, such as frugal skiplists \cite{kim2021rethink}, mitigate the retrieval problem, they incur substantial overhead for range scans. Second, deletions degrade performance because keys often remain in the B$^+$-tree even after they are logically deleted \cite{johnson1993performance}. Third, range scans can still conflict with concurrent insertions in highly contended pages, thereby reducing query performance. These concurrency problems with high contention might be mitigated by using lightweight latches for B$^+$-trees \cite{bottcher2020scalable, el2024lightweight}, which provide effective mechanisms for coping with high contention in conflict-prone workloads. 
% However, latch efficiency alone does not eliminate scalability concerns under mixed workloads. Fourth, advanced garbage collection also incurs substantial overhead, leading to undesirable runtime performance spikes. 

This paper takes a fundamentally different approach to version management and multiversion concurrency control. Instead of version lists and CoW approaches with path copying, we propose the concurrent multiversion B-tree (cMVBT), a concurrent index derived from the multiversion B-tree (MVBT) \cite{becker1996asymptotically}. The MVBT is a partially persistent generalization of the B$^+$-tree for version management with optimal asymptotic runtime and space performance. It can be viewed as a different type of CoW B$^+$-tree supporting latch-free queries, but avoiding path copying. Instead, space consumption scales linearly with the number of single-write transactions. In addition, range-scan performance asymptotically matches that of a B$^+$-tree that exclusively would keep records from the target version. Thus, the MVBT has already been recognized as a candidate for the concurrency problem \cite{kim2021rethink}, but the authors cautioned that using the structure should be done with care because “an upsurge of new versions ... result in undesirable structure modification, thus damaging OLAP queries.” However, this paper shows that the cMVBT (the proposed extension of the MVBT with concurrency support) can be used without concern. In fact, we provide a rationale for why the cMVBT is ideally suited to implementing MVCC. 

The advantages of the cMVBT can be summarized as follows. First, range-scan queries also run latch-free, and their performance is asymptotically optimal, independent of the mix of single-write transactions consisting of insertions, deletions, and updates. In general, deletions are directly supported with low overhead compared to other approaches using version lists. Garbage collection is not required to improve query performance, as is the case in \cite{bottcher2019scalable}.  Instead, on-demand garbage collection is enabled, seamlessly integrated with free-space management, resulting in substantially lower memory allocation counts. The cMVBT supports multiple write operations in parallel without batching. While our discussion in this paper is limited to using the cMVBT in main memory, all the techniques presented are applicable to external storage. In summary, our paper offers the following contributions:
\begin{enumerate}
% first figure in intro: Structure gives away a good access time to si queries.
% without cc, sk are good compared to ll, but: in-fact, mv still outperforms them.
% Missing: CC in MV.
    \item The paper presents the cMVBT, a redesign of the MVBT that introduces a novel multiversion concurrency control protocol based on optimistic latching and CaS. 
    %maybe assign a name for it;
    % Contribution is the design as a whole.
    \item The cMVBT enables latch-free range-scan queries and achieves high throughput for write operations. % TODO: rephrase
    % Inserts konstant; ~100 Inserts/s -> use a parm: Increase OLAPs/second. Response times remain perfect for MV; no conflicts with inserts, whereas the SkipLists show increased delay/conflicts.
    %Hint: Focus on the lock-free aspect, how to show this; No Conflicts -> reach high parallelism.
    \item Write operations and range-scans preserve the theoretical efficiency bounds established in the original MVBT work.
    % MV alone experiment: Showing standardized cost, e.g., like BTrees.
    % kinda proof: perhaps design aspect;
    % Critic from Reviewer: Overhead of the MV system. 
    \item An on-demand garbage collection mechanism is introduced, tightly integrated with free-space management.
    % MV: Experiment for showing effect, focus on mem stability.
    % one with GC and one without GC:
    % Memory overhead -> no interest;
    % *Only overhead* for GC measurements.
    % Only inserts and updates -> no in-place-updates.
    % for the update-in-place: Check plus; like a hybrid-structure/btree. must be explained from the concept itself.
    % CPU Overhead must be measured for GC. Must be minimal.
    % 3 Vars: GC + In-place; GC; No GC
    \item Experiments for mixed workloads derived from a standard benchmark demonstrate that the cMVBT significantly outperforms current approaches based on version lists.
    % ycbm; Check CH Benchmarks; Pi Benchmark is CRUD+S.
    
\end{enumerate}
% Next paper/Outlook: In-page GC.

The remainder of the paper is organized as follows.  Section~\ref{sec:relatedwork} reviews related work, discusses limitations of current multiversion data structures, and sketches briefly how they are addressed in our work. In addition, it introduces the MVBT. Section~\ref{sec:multiversiontree} discusses the new page layout for the cMVBT and explains how proactive reorganizations can be enabled without incurring asymptotic performance loss. Section~
\ref{sec:basicccalgo} presents the new concurrency control protocol for write operations using optimistic latching. Section~\ref{snapshot-query} discusses the algorithms for range queries on snapshots. 
Section~\ref{gc} introduces the on-demand garbage collection approach that is seamlessly integrated with free-space management. A few extensions of the cMVBT are discussed in Section~\ref{sec:extensions}. Section~\ref{eval} presents an experimental evaluation of the cMVBT in comparison with state-of-the-art approaches. Finally, Section~\ref{conclusion} concludes the paper.

%The remainder of the paper is structured as follows. 
%In Section~\ref{sec:relatedwork}, we present a review of related work in the area of multiversion data structures and concurrent systems, highlighting key approaches, their limitations, and the specific gaps our work aims to address. Section~\ref{sec:multiversiontree} introduces our concurrent multiversion B-tree (cMVBT). We first identify critical shortcomings of the original MVBT. This is followed by a detailed description of the new node layout and the new top-down reorganizations of the cMVBT. Then we present concurrency algorithms for OLTP and OLAP workloads that support latch-free range scans. Section~\ref{gc} introduces our garbage collection approach that is seamlessly integrated with free-space management. We further discuss extensions to the core design in Section~\ref{sec:extensions}. Section~\ref{eval} presents an experimental evaluation of the cMVBT, comparing it against state-of-the-art alternatives. Finally, Section~\ref{conclusion} concludes the paper.

\section{Related Work}\label{sec:relatedwork}

This section first presents an overview of related index structures designed for multiversion concurrency control and examines their limitations. We then introduce the original MVBT without any concurrency support and its four reorganization procedures.

Multiversion concurrency control (MVCC) has become a cornerstone of modern database systems \cite{stonebraker1986design,kemper2011hyper,diaconu2013hekaton,pavlo2016s}, enabling concurrent transaction processing while preserving isolation and consistency. Much of the research in MVCC has focused on transaction management \cite{van1996query, haapasalo2009transactions, dashti2016repairing, wu2017empirical,kim2022diva}, including high-level protocols and mechanisms to coordinate transactions and enforce consistency guarantees. However, ensuring efficient concurrent access at the index level introduces additional challenges.
%In contrast, our work addresses a complementary, orthogonal aspect: the design and concurrency control mechanisms within MVCC index structures.

Traditional index structures, such as B$^+$-trees, rely on in-place updates and therefore require synchronization mechanisms, such as latch coupling, to coordinate concurrent access by readers and writers \cite{bayer1977concurrency}. To reduce synchronization overhead, \cite{levandoski2013bw} introduces a latch-free B$^+$-tree using CAS (compare-and-swap) operations. While this approach performs well in read-dominated workloads with low update rates, it does not adequately address high-contention scenarios \cite{wang2018building}. In contrast, our method targets mixed workloads with insertions, deletions, and updates, allowing writers to proceed without interfering with readers and reducing contention even under heavy write pressure. Nonetheless, these existing approaches can still suffer from challenges such as operation contention, potential reader failures, and starvation under intense write workloads \cite{el2024lightweight}. Furthermore, ensuring consistency in such systems requires careful coordination, which introduces non-trivial overhead.

This paper addresses these limitations by introducing a redesigned tree structure tailored to MVCC systems. The key innovation is to design a concurrent append-only (copy-on-write) index derived from the MVBT \cite{becker1996asymptotically}. Our approach ensures that:
\begin{itemize}
    \item \textbf{Writers operate independently}, avoiding conflicts with concurrent reads. As a result, update operations do not interfere with committed data, which helps avoid latency spikes caused by contention.
    \item \textbf{Readers are latch-free}, operating on immutable committed data. Hence, reader queries (e.g., OLAPs) run without overhead and always succeed.
    \item \textbf{Reduced latch overhead in structural changes}, requiring only a single node latch for splits and at most two nodes for merges, whereas B-tree splits and merges involve multi-node coordination (parent, child, and sometimes siblings).
\end{itemize}

Our approach diverges from traditional MVCC index management with version chains \cite{wu2017empirical,kim2021rethink}, as the cMVBT logically stores versions in separate B-trees. Physically, these versions are merged into a DAG to guarantee linear space cost. 

Closely related to our approach are write-once indexes, such as the time-split B-tree \cite{lomet1989access} and copy-on-write indexes combined with path copying \cite{DBLP:books/daglib/0024051, Twigg:2011:SBV,sowell2012minuet,DBLP:journals/pvldb/0001BLP19}. However, the time-split B-tree does not support efficient deletions, whereas previous CoW approaches use path copying, which incurs high write amplification. The cMVBT overcomes these deficiencies as the amortized write cost of a single-write transaction is $\mathcal{O}(1)$. In addition, the cMVBT allows the adoption of concepts from concurrent B-trees \cite{el2024lightweight}, such as optimistic latching, which has been shown to be beneficial \cite{leis2019optimistic}. 

In addition, we also integrate garbage collection (GC) directly into the index structure with minimal overhead, providing efficient free-space management and ensuring that version pruning does not hinder performance. By combining these features, our design achieves consistency in multiversion index structures while enabling efficient  (OLAP) by isolating readers entirely from the effects of concurrent updates.

\subsection{Review of the MVBT}
The MVBT is a directed acyclic graph (DAG) that provides a compact physical representation of $N$ immutable B$^+$-trees, one for each version \cite{becker1996asymptotically}. Whenever an update occurs on the most recent B$^+$-tree, a new B$^+$-tree is logically created. This update model is known as partial persistence \cite{driscoll1986making}, and the MVBT is an asymptotically optimal persistent search tree. Unlike other approaches \cite{DBLP:journals/pvldb/0001BLP19,DBLP:books/daglib/0024051}, the MVBT avoids path copying, which would result in high write amplification, superlinear storage costs $\mathcal{O}(N \log N)$ when all $N$ versions are maintained, and consequently, high garbage collection overhead. Instead, the MVBT adopts the same design principles as B$^+$-trees, where updates typically affect only a leaf node and, only occasionally (e.g., when a node overflows), require additional nodes to be written. Therefore, for $N$ updates, the MVBT achieves $\mathcal{O}(N)$ space complexity and supports range scans at version $i$ with the same asymptotic complexity as a standard B$^+$-tree containing only the $i$-th version. 

To achieve these performance guarantees, data and index entries in the MVBT nodes additionally store their associated version information. In the original MVBT design, each entry is annotated with a validity interval $[vs, ve)$, indicating that the entry is live from version $vs$ up to $ve$, but not including $ve$. Entries corresponding to the latest version use an interval of the form $[vs, *)$. Additionally, the MVBT maintains multiple roots, with the versioned entries for these roots organized in a separate structure called $root^*$ (see Driscoll et al. \cite{driscoll1986making}). Because version intervals are disjoint and entries are appended in order, the entries in $root^*$ can be efficiently kept in version order.
%, as illustrated in \autoref{archmv}.

\begin{figure}[!htbp] % i think i will redesign this picture!
    \Description{MVBT Page Reorganizations visualized}
 \includegraphics[scale=0.45]{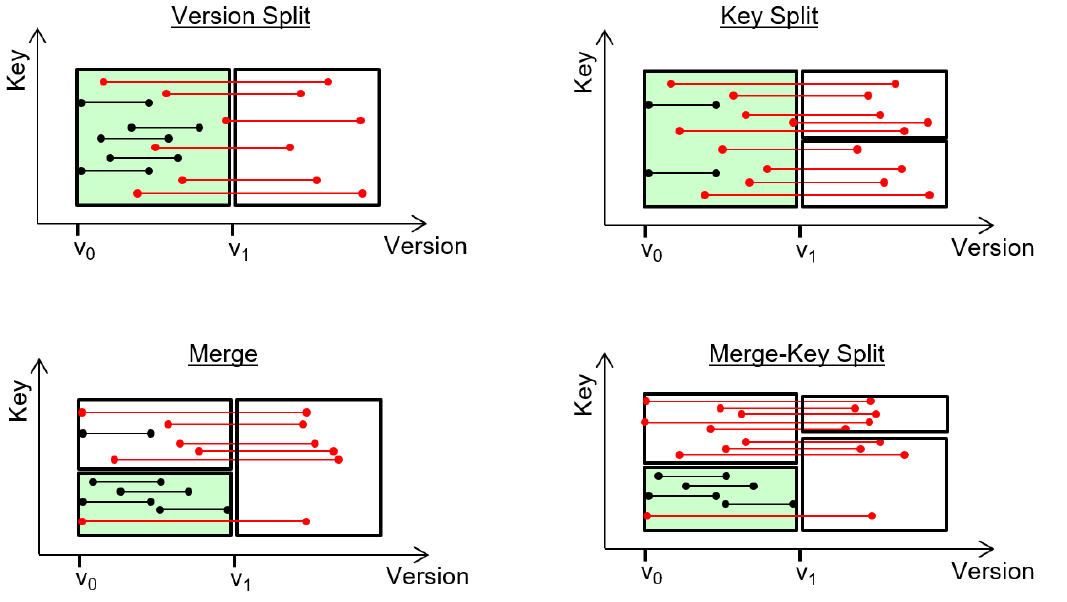}
	\Description{MVBT Reorganizations}
    \caption{The Four Page Reorganizations of the MVBT}
	\label{relatedwork:reorg}
\end{figure}

The linear storage space guarantee of the MVBT is based on two invariants. For that, let $B$ denote the page capacity (in number of entries/records). First, the \textit{weak-version condition} states that a linear fraction of the page capacity B belongs to the most recent version.  For simplicity, we will use $d = B/5$ as the minimum number of entries required for the most recent version. Second, the \textit{strong-version condition} states that after a node reorganization, at least $d$ operations on the node (insertions, deletions, updates) are required until the next reorganization can occur. These two invariants hold for all nodes except for the roots of the MVBT. For the roots, at least two entries must belong to the most recent version (assuming that more than $B$ records are in that version).  

As in standard B-trees, an operation starts at the root of the most recent version and traverses downward to a leaf. If the weak-version condition is violated or an overflow occurs, nodes are reorganized along the search path in reverse order. Below, we discuss the four reorganization operations on a leaf of the MVBT for $B = 10$ and $d = 2$, using the two-dimensional partitionings of the version-key space shown in \autoref{relatedwork:reorg}. Each leaf of the MVBT corresponds to a rectilinear rectangle, created at version $v_0$, with reorganization occurring at version $v_1$. The leaf that triggers the reorganization is emphasized (colored green). Versioned records are represented by black and red intervals, corresponding to being dead or alive at version v1, respectively. An insertion at version $v_1$ triggers a reorganization when the node contains more than B records. A reorganization always begins with a version split, in which live entries at version $v_1$ are copied from the original node to a new live node. If the strong-version condition is satisfied after this split, no further action is needed. Otherwise, additional reorganization steps are triggered. If the new node contains more than 8 ($= 4/5*B$) live entries, a key-split is performed first: entries are evenly distributed between two nodes using a split key from the key dimension, analogous to a split in a B+tree. If the node contains fewer than $B/5$ live entries, a merge with a live sibling is triggered. After a version split on the sibling node, the live entries from both nodes are merged into a final node, as shown in the lower-left plot of \autoref{relatedwork:reorg}. If the resulting node contains more than $4/5 \cdot B$ live entries, an additional key-split is performed, as illustrated in the lower-right plot of \autoref{relatedwork:reorg}. The two resulting live nodes then satisfy the strong-version condition. Hereafter, we use the term node reorganization to refer to version splits, merges, and key splits. Note that at most two new nodes are created during a reorganization.

\begin{figure}[!htbp]
\Description{An inner node structure visualization on a 2D-Coordinate system}
 \includegraphics[scale=.85]{{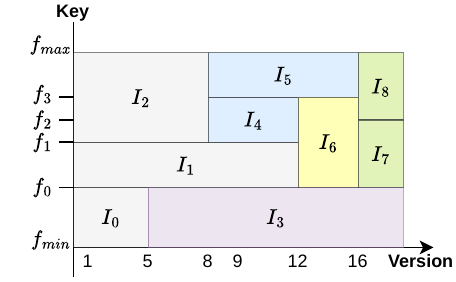}}
	\caption{2D Visualization of Entries in an Internal Node}
	\label{seqReorgs}
\end{figure}

% Muss noch nacheditiert werden
The example in \autoref{seqReorgs} illustrates a sequence of these reorganizations applied to nodes $I_0$, $I_1$, and $I_2$, which were alive at version 1. First, a version split occurs for node $I_0$ at version 5, creating node $I_3$. Next, a version-key split of $I_2$ at version 8 creates two nodes ($I_4, I_5$). At version 12, the nodes $I_1$ and $I_4$ are merged into node $I_6$. Finally, a merge-key split occurs at version 16. Nodes with equal colors \autoref{seqReorgs} indicate that these nodes are created at the same version, i.e., share the same birth time. The labels on the y-axis denote keys: $f_{min}$ and $f_{max}$ are the fence keys of the domain, whereas the others $f_0, \ldots, f_3$ are separators.

%
% Nur linke Seite, rechte Seite weglassen.
% Kann in einer anderen Fig später erläutert werden. 
%

\section{The Concurrent MVBT (cMVBT)}\label{sec:multiversiontree}
% This section presents the design of the cMVBT, an extension of the MVBT to support concurrent access. We discuss the changes of the underlying data structure and detail the concurrency algorithms for insertions, deletions, and updates.   

% ZZZ - need a rewrite after section is written
This section first presents our preliminaries and discusses limitations of the MVBT that have to be addressed in the design of the cMVBT. Then, we outline the new node layout and show how proactive reorganizations are enabled, before detailing the management of the roots. 

\subsection{Preliminaries and Motivating Deficiencies}
Before presenting the details of our current approach, we first summarize the underlying assumptions of the paper.
\begin{itemize}
\item Originally, the MVBT was designed as an external data structure that stores its data elements in nodes of capacity $B$ on external storage. Similar to the B$^+$-tree, it can also be used in main memory by setting $B$ accordingly, e.g., $B = 10$. Without loss of generality, this paper focuses on the cMVBT in main memory. However, all the MVCC techniques presented here are also applicable to the cMVBT on external storage. Note that this property does not hold for most other versioning approaches, e.g., those that employ version lists are only efficient in a main-memory setting. 

\item To simplify our discussions, we assume that data items are key-value pairs, where a key and a value are of fixed size. In addition, we assume that version numbers are integers with a constant size.  This is not a severe limitation, as the cMVBT can use the same techniques as the B$^+$-tree for variable-length data. We will discuss later how to treat variable-length keys and values in the MVBT. 

\item Finally, we emphasize that the paper focuses on the concurrent maintenance of the MVBT under insertions, deletions, updates and range scans for a visible version. Each operation is executed within its own transaction. Note that support for arbitrary ACID transactions is beyond the scope of this paper and will be addressed in our future work. 
\end{itemize}

%
% umgeschrieben
% 
The original MVBT has several shortcomings in supporting concurrent operations, which the cMVBT addresses. Each shortcoming below is accompanied by a brief idea of the corresponding cMVBT solution. 

\begin{itemize}
\item First, and perhaps surprisingly, there is no known concurrent extension of the MVBT, even though the original paper identified MVCC as a potential application. We will present a low-overhead concurrency algorithm for the cMVBT using proactive splits and optimistic latching that has recently proven highly effective \cite{leis2019optimistic}. 

\item Second, the MVBT associates a version interval with each entry to track its validity and significantly restructures the nodes during insertions or updates. In contrast, the cMVBT provides a true append-only representation, which is a prerequisite for its latch-free processing of range scans. 

\item Third, the MVBT requires explicitly specifying the snapshot (version) for a range scan. In contrast, the cMVBT automatically determines a fresh snapshot to guarantee latch-free query execution. 

\item Fourth, garbage collection was not thoroughly addressed in the original MVBT, as it assumed all versions must be retained to support queries on arbitrary versions. The cMVBT instead provides a low-overhead garbage collection mechanism to reduce space overhead. Separately, as is known for the MVBT, the cMVBT also offers (asymptotically) optimal range-scan performance regardless of the amount of garbage. 
\end{itemize}

\subsection{Node Layout of the cMVBT}
\begin{figure}[!htbp]
\Description{List of the entries of an internal node}
 \includegraphics[scale=.85]{{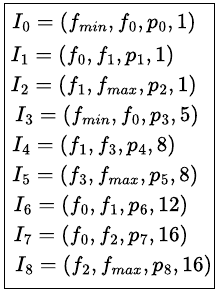}}
	\caption{List of Entries in the Running Example}
	\label{entriesinner}
\end{figure}

As discussed above, the cMVBT will offer a new node layout to ensure that updates are append-only. Instead of attaching version intervals to entries, an entry in the cMVBT uses only a version number. Consider the example in \autoref{seqReorgs} and assume that the entries of all nodes are kept in an internal node. Then, the corresponding entries are depicted in \autoref{entriesinner}. Every entry consists of the two fence keys of the node, the reference to the node, and the version number of its insertion, i.e.,  the snapshot of the operation that created the entry. 
Note that this results in slightly different node modifications. For an update, the cMVBT simply adds a new entry, whereas the original MVBT must also update the right boundary of a version interval. For example, the fourth entry $I_3$ is an update of the first entry $I_0$.
For deletions, we distinguish between internal nodes and leaves. For an internal node, there is no explicit deletion of an entry, but only updates can occur that cause an implicit deletion of previous entries. Consider, for example, the key-split of entry $I_2$ in \autoref{seqReorgs} that results in the creation of $I_4$ and $I_5$. These entries cover the key range of entry $I_2$, and thus, after their insertion, entry $I_2$ is not available in version 8 and higher. For a leaf, an entry is a triple consisting of the key, the value, and the version number. A deletion of a key is treated as an insertion of a tombstone entry that simply states that the previous key-value pair does not exist anymore from the given version. Note that the tombstone entries do not degrade efficiency because the cMVBT, like the original MVBT, guarantees the weak-version condition. Overall, each node modification creates at most two new entries appended to the end of the node, and the previous entries are not changed.

Important for concurrency is that every node maintains an 8-byte atomic value consisting of a latch bit, a counter for live entries, and a counter for dead entries. This value is used later in our optimistic latching approach to monitor the status of a node.

\begin{figure}[!htbp]
\Description{Representation of an internal Node's two sections, the committed (immutable) and uncommitted (mutable).}% old pic is sections1.pdf and .6 scale
 \includegraphics[scale=0.77]{{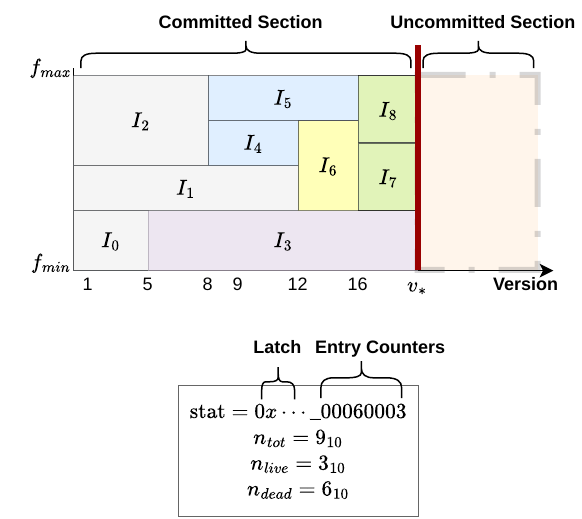}}
	\caption{Committed and Uncommitted Section of an Internal Node}
	\label{innersections}
\end{figure}

\autoref{innersections} depicts the node of our running example again.  
The new design divides each node into two distinct sections: the committed section, which contains immutable data, and the uncommitted section.  
The bold red vertical line separates the committed from the uncommitted section. The uncommitted section may contain partially written data by some writer, yet it is not visible to any other parallel thread and might be undone before committing. The data in the committed section can be read without latching, which is the foundation of our latch-free scan algorithm. The latch bit of the atomic value is set to 0, indicating that no writer is currently active on that node.

%The atomic method \texttt{Append\_and\_Commit} performs updates on a safe node as follows. First, it reads the atomic counter's value. Second, it appends the new data to the end of the committed version. Third, it updates the atomic counter according to the specific modification (insert, update, or delete) and unsets the latch bit.  Finally, the atomic counter is stored, thereby making the updates and the modified latch value visible.

%
%
%
\subsection{Proactive Reorganizations}
\label{sec:cmvbtinvariants}

The MVBT performs write operations (insertion, updates, and deletions) as in standard B-trees. An operation starts at the root of the most recent version and traverses downward to a leaf using the given key. If the weak-version condition is violated or an overflow occurs, nodes are reorganized along the search path in reverse order. In concurrent settings, bottom-up reorganization traversals and top-down write operations cause a high degree of conflicts. To avoid such conflicts, the cMVBT performs its reorganization steps proactively during the top-down traversal. This is a common approach used in B-trees and other search trees \cite{guibas1978dichromatic}. To introduce proactive reorganization steps for the cMVBT, we first introduce the notion of a safe node. 

\begin{definition}
Let $I$ be an internal node of the cMVBT that is alive. Let $n_{tot}$ be the total number of entries and $n_{live}$ be the number of live entries. $I$ is a safe node, if
\begin{enumerate}
    \item $n_{tot} < B-1$
    \item $I$ is the root and $n_{live} > 1$ or $I$ is not the root and $n_{live} > d$
\end{enumerate}
Otherwise, $I$ is an unsafe node. 
\end{definition}
The definition can be extended for a leaf node, except that a leaf is safe if $n_{tot} < B$ and $n_{live} > d$, i.e., it allows an insertion of a new record by insertion, deletion, or update.

The new write operation starts the traversal with the current root. For each node $I$ on the traversal path, the operation checks whether $I$ is unsafe. If so, the corresponding reorganization is performed. If the condition $n_{tot} \geq B-1$ for internal nodes or $n_{tot} = B$ for a leaf node is fulfilled, a reorganization is performed, as in the MVBT (version split, key split, merge-key split). All the nodes involved will be safe after the reorganization. The resulting new entries (referring to the newly created node) can then be inserted into the safe parent node. 
If $I$ is the root and $n_{live} = 1$, a new root is created, and a new entry is inserted into a list containing all the roots. We will discuss the management of the list in the next section. Otherwise, I is not the root and $n_{live} = d$. Then, a merge is performed, producing safe nodes again. The new entry of the merge can again be inserted into the safe parent node. In summary, by slightly adjusting parameter settings, the proactive splits and merges of the cMVBT are possible without affecting asymptotic performance, but with only a minor impact on storage utilization and height, because nodes are split and merged slightly earlier than necessary.

\subsection{Organization of the Roots}
As is known from the MVBT, the cMVBT can contain multiple roots, each valid over an associated time interval. While a write operation needs only access to the root of the live version, range-scan queries use a slightly older version that is visible to them. For that reason, the cMVBT retains the roots of the older version, though we expect they are not often used, as queries seek the freshest snapshot available to them.

\begin{figure}[!htbp]
\centering
\Description{A linked list for managing root*}
 \includegraphics[scale=.67]{{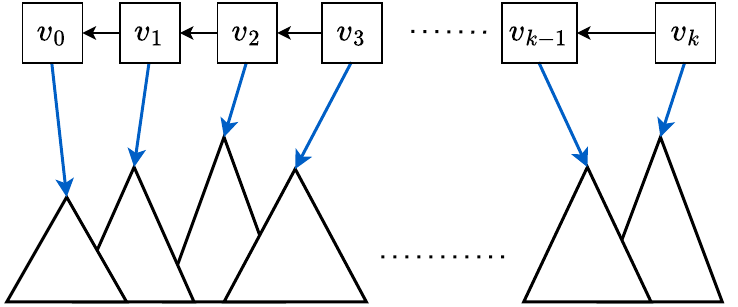}}
	\caption{Root$^*$ of the cMVBT}
	\label{archmv}
\end{figure}

Thus, a structure is required for managing the roots. The simplest approach is to use a linked (or double-linked) list as illustrated in \autoref{archmv}, where there are k+1 roots and k+1 entries in the list. 
As illustrated, the heights of two adjacent trees must be either equal or differ by 1.
The tree for the current root was created at version $v_{k}$ and remains valid until now. 
Obviously, such a simple list allows appending new entries fast, but it might be slow when a query requires access to a root referenced from an element at the end of the list. For that reason, we use a frugal skip list \cite{kim2021rethink}, which also provides constant-time access to the most recent tree root, whereas the access to older roots is logarithmic in the number of roots. 
% Reader operations thus traverse this root index from current to oldest, depending on the targeted snapshot (version). Writer operations always target root*, which is the first root in the list.

\section{Concurrent Operations}\label{sec:basicccalgo}
This section presents the MVCC algorithm
for the cMVBT that leverages optimistic latches and optimistic latch coupling for the write operation. The first subsection provides important preliminaries for the algorithms and introduces the concept of optimistic latches. The next subsection presents the concurrency algorithms for write operations, followed by a brief introduction to read operations.
% The final subsection details our latch-free query algorithm. 

\subsection{Preliminaries}
% Explain the global Counter
% Introduce Optimistic Latches and CAS
% Introduce the helper functions like
% XLatchOPt
In this subsection, we first introduce our version counter. We then present optimistic latching and the atomic compare-and-swap (CAS) operation in \autoref{alg:xlatchopt}. Last, we describe the function \textsc{Append\_and\_Commit} called by a write operation for inserting new entries into a node  \autoref{sec:write-operations}.

\subsubsection{Logical Clock for Version Ordering}
Multi-Version Concurrency Control (MVCC) systems employ a global logical clock to assign monotonically increasing timestamps to operations. This mechanism establishes a consistent ordering of operations and forms the basis for visibility decisions. In our approach, the global clock is implemented as an atomic counter, ensuring thread-safe, low-overhead timestamp allocation while preserving strict monotonicity. In the following algorithms, the function \textsc{NewVersion()} returns the next timestamp (version).

% \subsubsection{Non-blocking Latching}
% We employ optimistic locking based on atomic compare-and-swap (CAS) operations. Latches are non-blocking, and contention is resolved through retries rather than blocking. This approach is commonly used for short-lived critical sections due to its low synchronization overhead.

\subsubsection{Non-blocking Optimistic Latching}
\autoref{alg:xlatchopt} sketches the non-blocking latching of a node using a CAS-based approach, where the latch bit is encoded in the status value of the node. First, $stat$ is masked to remove the latch bit if set. This step is essential to ensure that the subsequent CAS operation (line 2) does not succeed based on a stale or already-latched state. In fact, it prevents multiple threads from acquiring the latch concurrently. The function \textsc{getState($I$)} returns the current status value for a given node $I$. In other words, if the CAS operation fails, it is because one of the following:
\begin{enumerate}
    \item The original status value, given as parameter in  \textsc{XLatchOpt}, is outdated and differs from the current one (\textsc{getStat($I$)}), even if no latch was present.
    \item Node $I$ is already latched. Thus, no other latch is allowed. Then, the actual status of I differs from the previous one given as an input parameter of \textsc{XLatchOpt}.
\end{enumerate}

Finally, our approach avoids the well-known ABA problem \cite{dechev2010understanding}. This is achieved because the status value includes counters that increment with every latch acquisition and release. As a result, the system never reuses the same status for different latches on the same node, thereby preventing ABA issues.

\begin{algorithm}[t]
\DontPrintSemicolon

\KwIn{Node $I$, Original Status $stat$ of Node $I$}
\KwOut{\textsc{Boolean}}

$stat \gets stat\, \&\, \neg write$\;

\If{\textsc{CAS}(\textsc{getStat($I$)}, stat, stat $\mid \text{write})$ fails}{
    \Return \textsc{false}\;
}

\Return \textsc{true}\;

\caption{XLatchOpt}\label{alg:xlatchopt}
\end{algorithm}
\subsubsection{\textsc{Append\_and\_Commit}}\label{sec:append-and-commit}
The steps for writing data into a node and committing the operation (\textsc{Append\_and\_Commit}) involve the following:
\begin{enumerate} % text from bernhard..
    \item Write into the beginning of the uncommitted section.
    \item Write the commit version into the header of the committed section.
\end{enumerate}
Thus, the uncommitted section may contain partially written data by some writer (by \textsc{Append\_and\_Commit}), yet it is not visible to any other thread and might be undone before committing. Consequently, readers can never access partially committed data, and there is no need to guard the access to the committed section. 

The observation is that the new data becomes physically accessible immediately after it is written, before the node's latch is released. Yet this doesn't mean readers can logically access it unless the visibility check permits it. The visibility is further discussed in \autoref{visibilitysection}.

\subsection{Write Operations}\label{sec:write-operations}
This subsection introduces the concurrent algorithm for insertions, deletions, and updates. 
\autoref{alg:write-traversal} (\textsc{WriteTraversal}) provides a sketch of the most essential steps given key $k$ and value $v$. For that matter, we first discuss \autoref{algo:currentroot} for computing the live root, followed by \autoref{algo:repair} for node reorganizations.
\begin{algorithm}[t]
\DontPrintSemicolon
\KwOut{Root node $r$, Root's $r$ status value $r_{stat}$}
\SetKwFunction{FGetRoot}{CurrentRoot}
\Repeat{$r$ is safe}{
    $(r,\, r_{stat},\, root^*_{stat}) \gets$ \textsc{getFirst($root^*$)}\;
    \If{$r$ is unsafe \textbf{and} \textsc{XLatchOpt}($root^*,\, root^*_{stat}$)}{
         $v\gets$\textsc{NewVersion()}\;
        \uIf{$r$ is in overflow}{
            $(r',\, r'_{stat}) \gets \textsc{SplitRoot}(r,\, v)$\;
        }
        \Else{
            $(r',\, r'_{stat}) \gets \textsc{PromoteChild}(r,\, v)$\;
        }
        \textsc{Append\_and\_Commit}$(root^*,\, r',\, v)$\;
        \textsc{UnLatch}($root^*$)\;
        % \textsc{UnLatch($r$)}\;
        $r \gets r'$\;
        $r_{stat} \gets r'_{stat}$\;
    }
}
\Return $(r,\, r_{stat})$

\caption{\textsc{CurrentRoot}}\label{algo:currentroot}
\end{algorithm}

\subsubsection{\textsc{CurrentRoot}}
 \autoref{algo:currentroot} repeatedly reads a root candidate via \textsc{getFirst($root^*$)}. This function also returns the candidate's status value, denoted by $r_{stat}$ and $root^*_{stat}$, the status value of $root^*$. If the observed root is already safe, it is returned immediately. Otherwise, the algorithm attempts to reorganize the root by acquiring a latch on $root^*$ using the optimistic latching procedure \textsc{XLatchOpt} (\autoref{alg:xlatchopt}).

Once the latch is successfully acquired (lines 3-4), a new version is generated via \textsc{NewVersion()}. Depending on the situation of the root node, either a root split (\textsc{SplitRoot}) is performed in case of an overflow, or the last remaining active child is promoted via \textsc{PromoteChild}, becoming the new root. In this case, the tree height decreases by one. The resulting node $r'$ becomes the new root candidate.

The update is then committed atomically using the function \textsc{Append\_and\_Commit} (\autoref{sec:append-and-commit}), which appends ($r',\,v$) to $root^*$. Finally, the latch is released, and the process repeats until a safe root is found.
\begin{algorithm}[t]
\DontPrintSemicolon

\KwIn{Node $p,$ Status value $p_{stat}$, Node $c$}
\KwOut{Node \textsc{$p$}, Node $p$'s status value $p_{stat}$}
 \If{\textsc{XLatchOpt}$(p,\,p_{stat})$ fails}{
        \Return \textsc{Exit\_and\_Restart()}
    }
    \uIf{$c$ is in overflow}{
    $v \gets \textsc{NewVersion}()$\;
    \uIf{$c$ requires version split}{
        $new\_entries \gets \textsc{VersionSplit}(c,\, v)$\;
    }
    \Else{
        $new\_entries \gets \textsc{KeySplit}(c,\, v)$\; 
    }
    \textsc{Append\_and\_Commit}$(p,\, new\_entries,\, v)$\;
    }
    \Else{
        $(s,\, s_{stat}) \gets \textsc{Sibling}(p,\ c)$\;
        \If{\textsc{XLatchOpt}$(s,\,s_{stat})$ fails}{
        \Return \textsc{Exit\_and\_Restart()}
        }
        $v \gets \textsc{NewVersion}()$\;
        \uIf{$c$ and $s$ can be merged}{
            $new\_entries \gets \textsc{Merge}(c,\ s,\ v)$\;
        }
        \Else{
            $new\_entries \gets \textsc{Merge-KeySplit}(c,\ s,\ v)$\;
        }
        \textsc{Append\_and\_Commit}$(p,\, new\_entries,\, v)$\;
        \textsc{UnLatch($s$)}\;
    }
    \textsc{UnLatch($p$)}\; % order should not matter
    $p_{stat} \gets \textsc{LoadState($p$)}$\;
    \Return $(p,\, p_{stat})$\;

\caption{\textsc{Repair}}\label{algo:repair}
\end{algorithm}

\subsubsection{\textsc{Repair}}
\autoref{algo:repair} lists the procedure \textsc{Repair} for maintaining the structural invariants in the tree under concurrent updates. It is invoked on a parent node $p$ and an unsafe child $c$. To restore the invariants, the algorithm applies a suitable reorganization on $c$.

%The algorithm establishes the invariants again by applying an appropriate reorganization of $c$. 
%that local violations, such as overflow or version underflow, are resolved while preserving correctness.

The algorithm first attempts to acquire optimistic exclusive latches on $p$ using \textsc{XLatchOpt}. The original status value $p_{stat}$ of $p$ is used to acquire the latch. Note that latching the parent node $p$ is sufficient because an unsafe node (like $c$) will not be modified anymore. If this latch fails, \textsc{Exit\_and\_Restart} is called that behaves similar as an exception. Then, Repair is killed, all acquired latches are freed, and the entire write operation is restarted (lines 1-2).
% Once the latch is acquired, a new version timestamp is generated via \textsc{NewVersion()} at line 3.

If there is an overflow of $c$, a new version is generated via \textsc{NewVersion()} (line 4) and the algorithm performs a suitable reorganization on $c$ (either a \textsc{VersionSplit} or a \textsc{KeySplit}). This returns a new set of entries that the function \textsc{Append\_And\_Commit} will promote to the parent node (line 9).

Otherwise, $c$ violates the weak-version condition. The method \textsc{Sibling} returns a sibling $s$ along with its status $s_{stat}$ (line 11). Then it attempts to latch the sibling accordingly (line 12). If the latch on the sibling fails, the operation is restarted (line 13). Otherwise, the algorithm either merges $c$ and $s$ using \textsc{Merge}, or redistributes entries between them using \textsc{Merge-KeySplit}. The resulting modifications together with version $v$ are promoted to the parent node using \textsc{Append\_and\_Commit} (line 19). Afterward, on line 20, the latch on the sibling node $s$ is released. Finally, the latch on the parent node $p$ is released (line 21). The new status of the parent node is read and returned together with the updated parent node $p$ (line 23).

\subsubsection{\textsc{WriteTraversal}}
In the following, we discuss \autoref{alg:write-traversal}.
First, the algorithm calls \textsc{CurrentRoot} to obtain the current root and the status value (line 1). 
It then traverses the tree top-down from the root to the target leaf with key $k$. To avoid conflicting bottom-up reorganizations, the algorithm performs proactive splits as introduced in section~\ref{sec:cmvbtinvariants}. It checks whether the child $c$ is safe (line 4).
%
% Haben wir schon vorher diskutiert. 
%
%, i.e., there is sufficient space for (at most two) new entries that are received from a potential reorganization of one of its children and that is upholds the weak underflow invariant, i.e, holds more than $d$ live entries. 
If not, the unsafe node is immediately repaired via the \textsc{Repair} function, and the traversal continues with the parent node $p$ (which \textsc{Repair} returns). 
%
% Haben wir davor schon erklärt. 
%
%Note that \textsc{Repair} performs a reorganization (split or merge) of the node and posts the new entries to the parent node (see lines 9 and 19 of the \textsc{Repair} algorithm).
After exiting the loop (line 10), $p$ refers to the target leaf that is optimistically latched via a call of \texttt{XLatchOpt}. Finally, the method \textsc{Append\_and\_Commit} applies the changes to the leaf using the new version $v$.
% We will provide the details in the next subsection (\autoref{sec:append-and-commit}). 

In analogy to classical latch coupling, the number of required latches is slightly lower. In case of an unsafe node $c$ being close to overflow, only one latch on its parent node is sufficient. A latch on $c$ is not required, since it can no longer be modified. Once the reorganization is complete, the latch on the parent node is released. For a merge, an additional latch on the sibling of $c$ is required. The reason is that the sibling is a safe node. Thus, we have to ensure it is not modified during the merge. All latches are released immediately after the corresponding modifications are complete, thereby minimizing contention.

\subsection{Read Operations}
Readers can safely access the immutable (committed) section of a node without acquiring any latch. This directly follows for historical nodes and nodes that trigger a reorganization, because these nodes can no longer be modified. If a reader accesses a live node, it first reads its status, which contains the two counters for live and dead entries (see \autoref{innersections}).
% These counters establish a \emph{happens-before} relationship with respect to concurrent modifications, ensuring that all committed entries are visible to the reader thread without additional synchronization.
In particular, a read operation can safely ignore the latch bit and thus traverse its node (i.e., the committed section) using the corresponding counters. 
% In summary, by combining on-demand node repair with optimistic latching, the algorithm ensures that writer operations can safely and efficiently modify the cMVBT while preserving its structural invariants.
This results in very fast and scalable reads, because:
\begin{enumerate}
    \item Readers never block or retry and are independent of workload contention.
    % \item Latching overhead is significantly reduced, because:
    % \begin{enumerate}
        % \item Latching logic is simplified.
        \item Writers only compete with other writers for the same data. There are no conflicts with read operations.
        % \item Proactive reorganization results in paths that never cascade upwards in modifications and does so by only requiring the minimal number of simultaneous latches, which further reduces conflicts.
    % \end{enumerate}
\end{enumerate}
These concurrency properties of read operations are essential for the concurrent processing of queries that will be detailed in the next section. 

\begin{algorithm}[t]
\SetAlgoLined
\DontPrintSemicolon
\KwIn{Key $k$, Value $val$}
% \KwOut{Exclusive-latched leaf node}
$(p,\, p_{stat}) \gets \textsc{CurrentRoot()}$\;
\While{$p$ is an internal node}{
    $c \gets \textsc{FindChild}(p,\,k)$\;
    \If{$c$ is unsafe}{
        $(p',\ p'_{stat}) \gets \textsc{Repair}(p,\, p_{stat},\, c)$\;
         $p_{stat} \gets p'_{stat}$\;
         $c \gets p'$\;
    }
    $p \gets c$\;
}
\If{\textsc{XLatchOpt}$(p,\, p_{stat})$ fails}{
    \Return \textsc{Exit\_and\_Restart()}
}
$v \gets $\textsc{NewVersion}()\;
\textsc{Append\_and\_Commit}$(p,\, k,\, val,\, v)$\;
\textsc{Unlatch($p$)}\;
\caption{\textsc{WriteTraversal}}\label{alg:write-traversal}
\end{algorithm}

\section{Snapshot Queries}
\label{snapshot-query}
In this section, we first outline how the cMVBT determines the freshest visible version (snapshot) for a given query in \autoref{visibilitysection}. It then follows a detailed discussion of point and range queries in \autoref{pointsandscans}.
\subsection{Snapshot Visibility}\label{visibilitysection}
So far, the cMVBT supports only single-row write transactions, each consisting of exactly one operation (insertion, deletion, or update). 
In addition, it supports read transactions (point queries and range scans) on arbitrary snapshots. 
%comprising multiple operations, yet we restrict ourselves to atomic transactions, where each transaction consists of exactly one operation (e.g., insertion or deletion). This makes OLTP workloads, which involve many fast-write transactions, suitable for our evaluation.
% \subsubsection*{Visibility Check}
However, in a transactional setting, a user does not specify the snapshot. Instead, the underlying system determines the valid snapshot, ensuring the query returns the most recent data. For that, every writer thread maintains a visibility watermark corresponding to its last local commit version, initially set to a sentinel value ($\infty$). When a writer transaction commits, it sets its visibility watermark to the commit version received from the global monotonic clock. 
When the reader transaction receives a query, it first computes the visible snapshot as the minimum watermark across all writer threads, and starts processing the query on that snapshot. This approach is well-established and used in other systems such as Steam \cite{bottcher2019scalable}, Hekaton \cite{diaconu2013hekaton}, and Peloton \cite{pavlo2017self}.

To avoid stalled versions caused by inactive or read-only threads, we reset a thread’s local visible commit version to a sentinel value whenever the thread aborts or transitions into a sequence of read-only transactions. During this phase, the thread does not participate in computing the globally visible watermark.
Upon switching back to updating transactions, the thread resumes publishing commit versions as before, thereby overwriting the sentinel value and re-participating in determining the globally visible commit version.

% To determine the set of fully committed versions that are safe for snapshot reads, reader transactions compute the minimum across all thread-local visible commit versions. This minimum defines a global visibility frontier such that all versions with commit timestamps less than or equal to it are guaranteed to be committed and visible to all threads.
% This approach is well-established and used in systems such as Steam \cite{bottcher2019scalable}, Hekaton \cite{diaconu2013hekaton}, and Peloton \cite{pavlo2017self}.

\subsection{Point Queries and Range Scans}\label{pointsandscans}
This section presents the algorithms for point and range queries on a given version $v$. The first step for both types of queries is to determine the root, which is the entry point for version $v$. For that, we traverse root$^*$ (the list of all roots) starting from the most recent entry until an entry is found whose version is lower than or equal to $v$.
% As the length of root$^*$ is expected to be short, a simple linked list suffices for managing the entries. If not, a frugal skip list could be used to guarantee logarithmic worst-case cost. 

\begin{figure}[!htbp]
\Description{Point querying on an inner node visualized via a 2D-Coordinate system.}
 \includegraphics[scale=.9]{{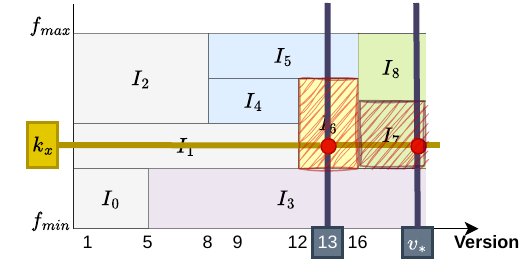}}
	\caption{Two Point Queries for an Internal Node}
	\label{pointquery}
\end{figure}

Let us first consider the two point queries illustrated as dots in \autoref{pointquery} where the entries refer to our running example. The one query uses key $k_x$ in version $v_*$, and the other uses key $k_x$ in version 13. For the query on version $v_*$, it suffices to examine the live entry $I_7 = (f_0,f_2,p7,16)$ because $16 < v_x$ and $f_0 \leq k_x \leq f_2$. Thus, the query continues with the child $p_7$ that $I_7$ refers to. For the second query with version 13, entry $I_7$ is again visited, but does not qualify. Thus, the remaining entries are traversed until the first match is found. In this case, it is entry $I_6$. Overall, the worst-case cost is linear in the number of elements in the node. 

%We demonstrate how reader operations access committed data within a node. For that matter, we first show how a point query is processed, then proceed to a range query.
%Consider the following \autoref{pointquery}, where we point query for key $k=k_x$, once at time $v=13$ and another at time $v=v_*$.
%The two small red circles are the query intersections with the node entries. The matched entries (squares) are those that are ultimately traversed to reach a leaf node. At a leaf node, the record data is first filtered by version (since the data is sorted by version), then by exact key match. The root node itself, where the query is started, is determined only by the query version, since all roots cover the entire key space, meaning the frugal list is iterated from newest to oldest roots and halted once a root is found with $root_v \le v$. Queries targeting $v_*$ always traverse $root^*$ directly. A root fulfilling that condition is always present, as the first root starts with $v=0$, the smallest valid version. Querying data with $v \le 0$ is invalid, as it's not meaningful to query a clearly empty index.

% Moreover, the first query from \autoref{pointquery} with $v=13$ (and any $v \in[12,\, 16)$) selects the entry $I_6$ to continue the traversal. The other point query with $v=v_*$ (and any $v\in [16,\,  v_*+1)$) traverses the entry $I_7$.

\begin{figure}[!htbp]
\Description{Range query execution example showing qualifying nodes via a 2D-Coordinate system.}
 \includegraphics[scale=.9]{{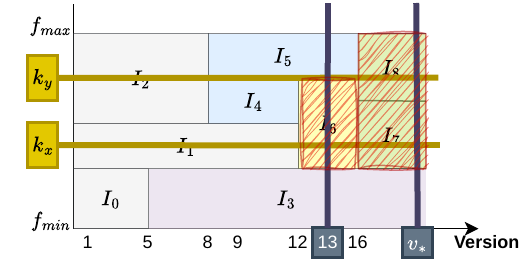}}
	\caption{Two Range Queries for an Internal Node}
	\label{rangequery}
\end{figure}

For a range query with key range $[k_x,k_y]$ and version $v$, the processing of a node is more complicated because multiple entries can qualify. In the following, we present two approaches: an iterative scan algorithm and a sweep-line algorithm. The iterative scan algorithm initiates multiple point traversals. Let $f_{min}$ and $f_{max}$ be the fence keys of the node. The first search starts with key k = max($f_{min}$, $k_x$). Let $I$ be the qualifying entry and $f^I_{min}$, $f^I_{max}$ the fence keys of $I$. While $f^I_{max} < min(k_y,\,f_{max})$, the search continues with the next key $k = f^I_{max}$. Consider the range query $[k_x,\,k_y]$ at version $v_*$ in \autoref{rangequery}. The first point query starts with key $k_x$ and returns entry  $I_7$. The next and final point query uses key $f_2$ and returns $I_8$. The algorithm terminates because the upper fence key of $I_8$ exceeds $k_y$. Overall, the worst-case performance is quadratic in $B$ (the capacity of the node), but as the number of qualifying entries is often observed to be ($\sqrt B$) \cite{arge1998scalable}, the practical runtime is O($B \cdot \sqrt B$). In addition, the repeated scans are cache-friendly, so only the first is costly.

A different approach moves a sweep-line from right to left over the temporarily ordered sequence of entries in the node. In the sweep-line status, we maintain the current coverage of the range $[max(f_{min},\,k_x),\, min(f_{max},\,k_y)]$. The sweep-line moves until it rea-ches the state that contains all entries matching the key range at version $v$. In our example in \autoref{rangequery}, the initial sweep-line status for the range query at version 13 contains the entries $I_7$ and $I_8$. However, the next entry, $I_6$, covers the entire query range at version 13. $I_6$ is the only result returned by the sweep-line method. The worst-case complexity of the method is $\mathcal{O}(B \cdot \log B$), but we observed that its practical performance for small $B$ is not better than that of the iterative scan algorithm. Thus, we decided to use the iterative scan algorithm in our implementation.

Overall, the range scan algorithm performs a depth-first traversal of the tree that consists of the nodes that belong to the required version $v$ and applies the iterative scan algorithm to every visited internal node. 

%$in a depth-first-search-like fashion, incrementally advancing the search key range, following the approach mentioned in \cite{haapasalo2009transactions}.

%For example, consider the same two point queries from the previous example, but now instead of querying $k=k_x$, we query the key-range $k\in [k_x,\, k_y]$ and visualize it on \autoref{rangequery}:
%For the first query targeting version $v=13$, the entry $I_6$ is selected because it's the only entity that spans the key range $[k_x,\, k_y]$ in that version. Thus, that query traverses that entry. It is noted that no more entries are traversed, because the search-key interval is exhausted by the first entry, namely, $I_6$.

% The other range query at $v=v_*$, on the other hand, intersects two entries, $I_7$ and $I_8$, because both of these blocks intersect the query's key-range $[k_x,k_y]$ at that version.
%Furthermore, in the second query, if $v=16$, the same entities would qualify. In other words, a range query selects entries as long as the key range of the entry qualifies and its version remains less than or equal to the target version of the query. If matches were found (i.e., in a leaf node), partial results are accumulated. Then, the query continues by adjusting its key range accordingly (advancing $ k_x$ to $ f_{max} +1$). The query thus ends when its key range is empty (i.e., when $k_x > k_y)$. This method of executing range queries has the additional benefit of returning results that are already partially sorted by key. Additionally, the query can be partially executed, e.g., depending on the desired number of results.

\section{Garbage Collection}\label{gc}
Garbage collection (GC) is essential for MVCC systems with version chains, as the length of these chains significantly impacts query performance, as discussed in \cite{bottcher2019scalable}. Common approaches perform GC in epochs, causing activity spikes, or run GC as an additional task at query time.  Recall, however, that the number of maintained versions does not impact the practical query performance of the cMVBT. The only reason for GC is to reclaim space early to avoid memory overflows. This less demanding task gives the cMVBT greater flexibility in addressing the GC problem. In the following, we first introduce our on-demand approach for GC. Then, we will discuss an optimization for overwriting entries within nodes.

%Garbage collection is required in MVCC systems because OLTP workloads continuously produce new versions, leading to version chains and index entries that grow without bound. Without a reclamation mechanism, memory consumption increases indefinitely, and lookup performance degrades as chains become longer. To address this, MVCC engines integrate dedicated garbage collectors that periodically reclaim versions no longer visible to any active transaction. Several design approaches exist, each with different trade-offs. Epoch-based garbage collection is widely used in modern in-memory systems due to its simplicity, scalability, and low runtime overhead, relying on coarse global epochs to determine when old versions are safe to free.

%Cooperative (COOP) garbage collection distributes reclamation work among worker threads and often leverages fine-grained visibility checks to prune version chains more aggressively. Reference-counting–based GC tracks per-version references and frees versions immediately after all readers have left the version, offering precise reclamation at the cost of significant synchronization overhead.

%Other approaches, including background vacuuming or hazard-pointer–based memory reclamation, exist but are less common in high-performance in-memory MVCC engines due to higher overheads.
\subsection{On-demand GC of Nodes}

The base method of the cMVBT is closely related to the one originally described for the MVBT. The idea is that nodes triggering a reorganization will become dead. These dead nodes are stored in an ordered list by death time. In the following, we refer to it as the graveyard list. Thus, the oldest nodes are in the front, and the nodes that recently died are at the end of the graveyard list. In addition, the cMVBT manages the running queries in an ordered list by start time. This list is called the active query list. Whenever a query finishes, it is removed from the active query list. If the query is the oldest, the front of the graveyard list is checked for nodes that are older than the currently active oldest query. All of these nodes can be released, thereby freeing up their storage space. This approach is also known from the methods using version chains. This could lead to a performance spike, as many nodes in the graveyard list may be affected.

By contrast, the cMVBT uses a different on-demand GC strategy, in which writer operations apply small GC steps without incurring overhead.  Whenever a writer operation triggers a reorganization, the new nodes are generally allocated via the underlying memory system (which causes additional cost). However, the cMVBT uses its graveyard list as a free list. Whenever the cMVBT requests a new node, the oldest node in the graveyard index is first checked for reuse. Thus, at most two nodes are removed from the graveyard list at a time. This is a valid strategy in case the number of elements in the newest version remains stable or increases. Only if that is not the case (which is quite unlikely), we switch to a more aggressive strategy such that an ordinary write operation (without causing a reorganization) could free space in the graveyard list. Ideally, this strategy leads to a substantial reduction in interaction with the memory system, such that many page allocations are served from the graveyard list.

\subsection{Updates In-Place}
Versioning is only necessary while queries are being issued. If not, it would also be possible to allow overwriting committed data records and index entries in the cMVBT. In the extreme case, the cMVBT would behave like a B$^+$-tree. Consider a situation in which a key $K$ is valid from version $v_0$. If there is an update to version $v_1$ on $K$ and there is no active query between $v_0$ and $v_1$, it is safe to overwrite the old version of $K$. However, such an update-in-place requires more sophisticated synchronization. This is beyond the scope of this paper, and we will leave details for our future work. 

\section{Extensions}\label{sec:extensions}
The cMVBT can also support common B$^+$-tree extensions [20] with a moderate amount of changes, e.g., variable-length records and compression. Consider, for example, records with variable-length values. Then, we could partition a node into two parts, where the first serves as a storage area for variable-length values and the second manages the fixed part of the records (with offset and length information for the values). The versioning mechanisms operate on the fixed part of the records, ensuring that updates and versioning remain consistent without affecting the dynamic layout of the values.
Despite the challenges posed by variable-length entries, node
splits, and repairs, the cMVBT remains feasible. When a node overflows, the operation can compute the space required for the separator key and retain this information while checking whether
the parent has sufficient space to accommodate it. If the parent
cannot accommodate the new separator, it can be split proactively. This approach ensures that the standard cMVBT splitting and reorganization logic applies, even when the values are of variable size.

Techniques such as prefix compression or key delta encoding can
reduce node size and improve cache efficiency. These methods are
directly applicable to the MVBT, just as in traditional B$^+$-trees,
because compression operates on the node metadata and does not
interfere with the versioning mechanism. For example, it would be possible to store the common prefix of the fence keys in the header of a node and to save this prefix for every record stored in the node. As a result, nodes can store more entries per page, improving both memory utilization
and I/O efficiency.

However, the focus of this paper is not on these aspects, and we leave their concrete effects on the cMVBT for our future work.

\section{Experimental Evaluation}\label{eval}
This section describes a preliminary set of experiments and discusses the most important achieved results. For that,
we differentiate between results showing latency and throughput
in a comparative study of the cMVBT with list-based approaches. Finally, we will discuss specific performance features of the cMVBT. 

% In our experiments, we use the standardized YCSB benchmark \cite{cooper2010benchmarking} and its workload specification methods to configure a range of workloads with different access patterns.
% Furthermore, for GC-enabled workloads, the database system maintains a concurrent auxiliary skip list to track the targeted snapshot for all readers. This mechanism thus tracks all snapshots in use at any time, since each writer transaction accesses only the current snapshot.
 \subsection{Setup}
Our experiments are performed on a system equipped with an
AMD EPYC 7742 64-core processor and 512 GB of main memory,
running Ubuntu 22.04.1. All our implementations, including the
one for the cMVBT, are written in Rust. All trees are created in
main memory. In addition to the cMVBT, we consider a concurrent
B$^+$-tree \cite{el2024lightweight} and its common extensions that support versioning.
The one extension employs linked lists called Version Chains, while
the other uses frugal skip lists as implemented in vWeaver [16]. The page size used by all structures is 4KB, resulting in page capacities
$B$ = 125 and $B$ = 170 for the cMVBT and B$^+$-tree, respectively. 
% Finally, we added an append-only latch type, specifically designed to allow concurrent readers alongside a writer, provided that the writer is appending data only, i.e., extending the version chain rather than physically moving or removing data. This enables the B$^+$-tree to operate fully in multiversion via the version chains.
% The source code for the Multiversion B$^+$-tree via version chains is found here: \href{https://github.com/umr-dbs/BTree-MVCC-Version-Chains}{github.com/umr-dbs/BTree-MVCC-Version-Chains}

Our data sets and workloads are generated using the standardized YCSB benchmark \cite{cooper2010benchmarking}. Our default data distribution is uniform, but some of the experiments also use Zipf distributions with parameter $\alpha$.
Our OLTP workloads consist of a mix of insertions, updates, and deletions. In an initial build-up phase, the first 10K operations are only insertions by default. In the second phase, we start applying a mix of updates, insertions, and deletions. A parameter, the so-called update rate, specifies the percentage of updates, whereas the number of insertions is kept equal to the number of deletions. Thus, the number of records in the most recent version will not change during the second phase. It is important to note that our experimental study considers deletions that have rarely been examined in previous studies on multiversion concurrency control. 
Our OLAP workload consists of queries returning the entire dataset for a given version. HTAP workloads consist of a mix of writer threads (performing insertions, updates, and deletions) and reader threads running scans. In our experiments, the default number of writer and reader threads is 32 and 16, respectively. When a thread has finished its task, it immediately receives the next one. 

\subsection{Results}
%For scans, we select all live keys in a committed snapshot. Additionally, we uniformly query all snapshots.

\begin{figure}[b]
\Description{Plot showing scan latency in different OLTP workloads.}
 \includegraphics[scale=.9]{{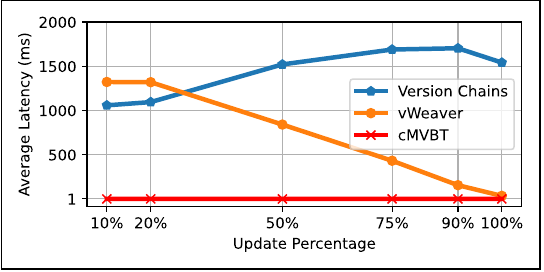}}
	\caption{Scan Latency as Function of the Update Percentage}
	\label{updatesvslatency}
\end{figure}

\subsubsection{Scan Latency}
In our first set of experiments, we evaluate the scan performance of the three multiversion methods under OLTP workloads with varying update percentages (10\%, 20\%, 50\%, 75\%, 90\%, 100\%). For that, all the snapshots created by the OLTP workload are kept (no garbage collection). After processing the workload, 100K scan queries are uniformly distributed across all versions and executed sequentially. 

\autoref{updatesvslatency} depicts the average scan latency as a function of the update percentage. 
The 100\% update workload is comparable to that in the experimental setup of \cite{kim2021rethink}. For this setting, our results confirm the superiority of vWeaver over Version Chains that degrade under mixed workloads for an increasing update rate and
exhibit a slight slowdown-stop for update rates close to 100\%.
\autoref{updatesvslatency} clearly depicts a cross-point between vWeaver and Version Chains. vWeaver is slower than Version Chains for workloads with few updates, but becomes comparable as update rates increase, and eventually superior. The reason is that version chains are very short at low update rates, so the overhead of the advanced list organization in vWeaver degrades performance. For higher 
update rates, this no longer has a noticeable impact. 

Overall, the cMVBT is the clear winner. It is already slightly superior to vWeaver in the updates-only case, which is indeed the best case for vWeaver, as deletions do not occur. Moreover, the performance of the cMVBT is constant for all update rates and is substantially faster than Version Chains and vWeaver. These results also confirm the theoretical findings of the cMVBT that the cost of a snapshot query is determined by the result size (which is constant for all versions in this experiment) and is independent of the mix of operations in the OLTP workload. In particular, the cMVBT can handle deletions very well. 

Furthermore, the results clearly show that Version Chains and vWeaver suffer from workloads with deletions because all data items accumulate in the underlying B$^+$-tree. In particular, deletions cause problems in vWeaver that introduce links between adjacent skip lists. Note that the number of links between adjacent skip lists can be high, and thus, the deletion of entire lists becomes inefficient. Thus, we decided not to remove lists, with the effect that scan queries must access lists unrelated to the query. Because weaving adjacent lists only gives a benefit for very long version lists and very long lists rarely occur in our experiment (except in this one), we will not consider the weaving technique in the next experiments, but use only the pure frugal skip lists \cite{kim2021rethink}. 
% Queries become slower the more data are and increasingly so as data volume . 
%This means that scan queries still traverse many keys because no version is valid for the queried snapshot, thereby dramatically increasing latency.
% Compared with our cMVBT, latency remains low, except during an initial warm-up phase. The warm-up phase is the period during which the cMVBT evolves from its initial, partially versioned state to a statistically stable structure under a stationary workload. In the steady state, structural changes are limited to localized version splits, while the overall shape of the tree remains stable, resulting in predictable, consistently low scan costs.

\subsubsection{Concurrent Throughput}

In our next experiment, we considered a concurrent HTAP workload with 32 writer threads and 16 reader threads. Instead of latency, we report throughput, which is more commonly used in concurrent evaluations. The writer threads used the same settings as in our previous experiments: an initial phase of 10K insertions followed by workloads of 1M operations with varying update rates. The 16 reader threads run snapshot scans on the freshest visible version as outlined in \autoref{visibilitysection}. All reader threads were terminated immediately upon completion of all the write operations. 

\begin{figure}[t]
   \Description{Throughput of scans and OLTP operations across different update percentages in the workloads without GC.}
  \includegraphics[scale=.85]{{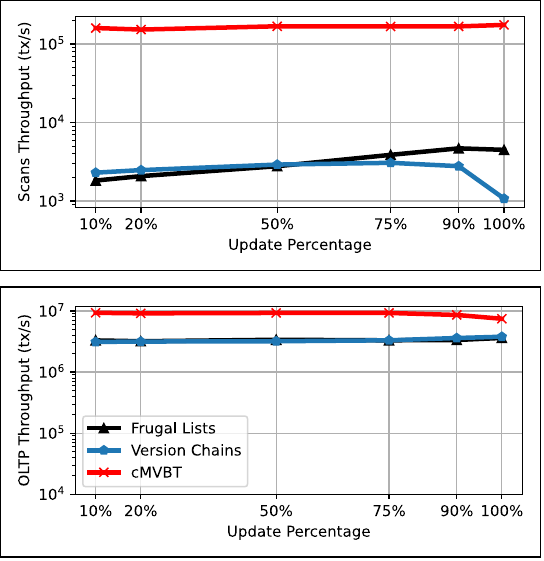}}
	\caption{Throughput in Concurrent OLTP Workloads Without GC}
	\label{concurrentoltp}
\end{figure}

\autoref{concurrentoltp} provides two results plots for this setup. The upper one shows the throughput in snapshot scans per second, and the lower one the throughput in OLTP operations per second. As mentioned above, vWeaver uses pure frugal skip lists, but no links among adjacent lists. 
vWeaver exhibits slightly lower OLTP throughput than Version Chains due to additional overhead from managing version lists. Moreover, there is no advantage of vWeaver for scan throughput, because queries run on fresh snapshots that often require only the first elements in a list.
The cMVBT consistently outperforms both vWeaver and Version Chains in OLTP throughput and scan throughput.
The reason for its high scan throughput is its latch-free query processing. The performance advantage in OLTP throughput is due to the cMVBT having no conflicts with reader threads.

% Although both vWeaver and version chains require traversing multiple versions during scans, their performance differs due to the stability of the underlying link structure. In version chains, pointers are largely append-only and remain stable during traversal. In contrast, vWeaver introduces cross-node version links through ongoing updates, increasing structural indirection and traversal complexity under concurrent OLTP workloads. This leads to higher cache miss rates and reduced scan throughput.
% Compared with the previous experiment in \autoref{updatesvslatency}, this experiment reports the real-world costs of transactions rather than a purely isolated measurement (ideal scenario).

\begin{figure}[b]
   \Description{Throughput of scans and OLTP operations across different update percentages in the workloads with GC enabled.}
  \includegraphics[scale=.85]{{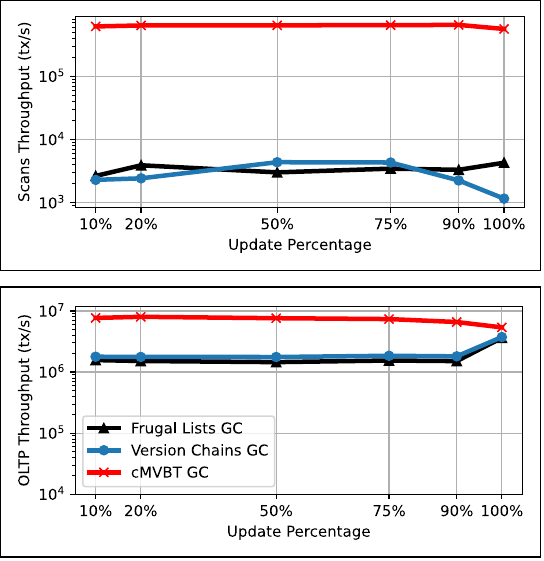}}
	\caption{Throughput in Concurrent OLTP Workloads with GC}
	\label{concurrentoltpgc}
\end{figure}

We also conducted the same experiment with GC enabled, as shown in \autoref{concurrentoltpgc}. We equipped the B-tree with a garbage-collection technique that runs during reorganizations, i.e., page splits and merges remove deleted keys, ultimately preventing unsustained growth.
Moreover, consistent with the results in \autoref{concurrentoltpgc} and \autoref{concurrentoltp}, both Version Chains and vWeaver remain slower than cMVBT in scan and OLTP throughput, regardless of whether garbage collection (GC) is enabled or not. This suggests that the underlying data structure is a dominant contributor to query throughput.
In contrast, cMVBT organizes data by version into contiguous blocks. Because readers access only immutable data, synchronization overhead is low, and the block-oriented layout yields significantly improved cache efficiency.

In addition, the primary role of GC in the cMVBT is to maintain stable memory consumption by retaining only relevant data and preventing unbounded memory growth. The system continuously reclaims obsolete versions, ensuring that memory usage remains sustainable even under long-running workloads. Moreover, reclaimed nodes are recycled internally, which reduces the frequency of memory allocation and deallocation. This reuse not only lowers memory management overhead but also improves cache behavior and overall execution efficiency.

\subsubsection{Scalability of Throughput}
\begin{figure}[b] % !htbp
\centering
\Description{Figure displaying the performance of the cMVBT vs. Frugal Lists in different concurrency levels}
 \includegraphics[scale=0.89]{{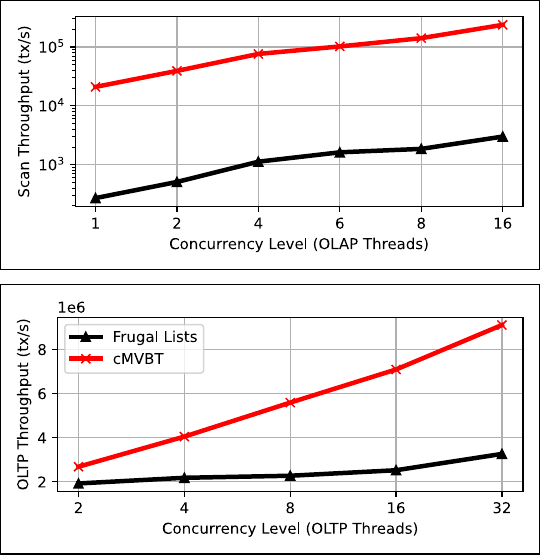}}
	\caption{Scalability of Throughput}
   \label{fig:mvbtcc}
\end{figure}
Next, we examined the scalability of our concurrency control strategy as a function of the number of threads for both OLTP and OLAP operations in an HTAP setting (60\% update rate). \autoref{fig:mvbtcc} displays the results for the cMVBT and B$^+$-tree with frugal skiplists. OLTP throughput grows rapidly for the cMVBT, primarily because there are no conflicts with concurrently running scan operations. In general, root-to-leaf traversals are latch-free and require no validation of optimistic latches. Only leaf nodes are briefly latched during writes. To the contrary, OLTP throughput does not scale well for frugal lists, although the B$^+$-tree also employs optimistic latching. However, the crucial difference to the cMVBT is that the optimistic latches in a B$^+$-tree require a validation process \cite{leis2019optimistic} with repeated atomic version checks (e.g., acquire load that ensures memory operation ordering). This synchronization overhead limits the scalability under mixed OLTP/OLAP contention. 
Read operations, by contrast, scale without introducing contention in the cMVBT but are bounded by memory bandwidth. Frugal lists, on the other hand, show a similar scaling trend for OLAP workloads, but additional latch contention reduces the throughput. Overall, the throughput is two orders of magnitude lower than that of cMVBT, independent of the number of threads. This is consistent with the results of our previous experiments. For that reason, we did not report results for version chains that perform similarly to those of frugal skiplists.

\subsubsection{Specific Features of the cMVBT}
This section provides important insights about the cMVBT. 

\begin{figure}[t]
\centering
\Description{Figure displaying the percentages of retries for various Zipf distributions}
 \includegraphics[scale=0.89]{{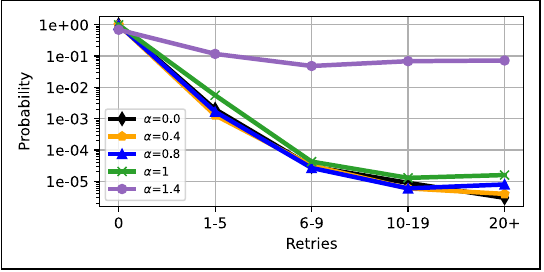}}
	\caption{Probability of retry groups for various Zipf distributions}
   \label{retries}
\end{figure}

First, we examined the overhead of optimistic latches in an OLTP experiment with 1M insertions. The results in \autoref{retries} show the probability of retries (logarithmic scale) as a function of five retry groups (0, 1-5, 6-9, 10-19, 20+) for different kinds of Zipf distributions ($\alpha = 0, 0.4, \ldots, 1.4$). For example, retry groups 0 and 1-5 mean that no retries and 1 to 5 retries are performed, respectively, until the operation succeeds. The graphs show that the probability of a retry is low, except for $\alpha = 1.4$. In general, the cMVBT can cope with highly skewed data. Other studies on optimistic latching \cite{ding2018improving} have not examined higher values for $\alpha$ as in our experiments. 

\begin{figure}[!htbp]
\centering
\Description{Figure displaying nodes reuse vs. node allocations in an OLTP workload}
 \includegraphics[scale=0.89]{{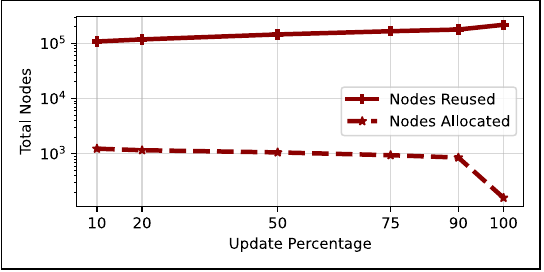}}
	\caption{Nodes Reuse vs. Nodes Allocations in an OLTP Workload via GC}
   \label{fig:reuse}
\end{figure}

Second, we examined the value of the adaptive GC. \autoref{fig:reuse} displays the number of node allocations in the MVBT where the upper curve shows the number of nodes reused and the lower curve the nodes that are allocated from the memory manager. Results are again given as a function of the update rate. Overall, the results show that nearly all node allocation are taken from the GC list, whereas only 1\% are from the memory manager. In case of updates only, the MVBT only applies version splits, and thus, the reuse is almost 100\%. One reason for this result are also due to our workloads that keep the number of live records very stable over time. 
% MVBT vs. BTree without versioning i would omit;
% also, now the MVBT is significantly faster due to
% minimized conflicts via requiring fewer latches

% Fourth, we measured the OLTP performance of the cMVBT against a B-tree that maintains only the most recent version (i.e., without versioning support). Overall, the throughput of a B-tree is about 50\% higher than that of the cMVBT, as shown in other performance studies of multiversion B-trees \cite{becker1996asymptotically,varman2002efficient}. But it experiences a complete performance collapse as concurrent writers increase (e.g., here 32 threads), as shown in \autoref{fig:mvbtvsbtree}.
% % At first glance, this may seem high, but it is consistent with other performance studies of multiversion B-trees \cite{becker1996asymptotically,varman2002efficient}.

% \begin{figure}[!htbp]
% \centering
% \Description{Figure displaying the performance of the cMVBT vs. B+Tree without versioning}
%  % \includegraphics[scale=0.89]{{figs/mvbtvsbtree.pdf}}
%  % This experiment for BTree is contention bound, not memory latency bound. TODO: Check
%  \includegraphics[scale=0.89]{{figs/mvbtvsbtreeOpt.pdf}}
% 	\caption{cMVBT vs. BTree without Versioning in different OLTP Workloads}
%    \label{fig:mvbtvsbtree}
% \end{figure}

% \begin{figure}[!htbp]
% \centering
% \Description{Figure displaying the percentages of retries for various Zipf distributions}
%  \includegraphics[scale=0.89]{{figs/skew1.pdf}}
% 	\caption{Probability of retry groups for various Zipf distributions}
%    \label{retries}
% \end{figure}

% \newpage
\section{Conclusion and Future Work}\label{conclusion}
This paper studied the concurrency problem for an HTAP workload with range scans and write operations, each of which consists of a single operation (insertion, deletion, and update), also known as single-row write transactions. To address this problem, we presented the cMVBT, a concurrent multiversion index structure designed to deliver high performance for these HTAP workloads. Rather than using version chains, the cMVBT employs an extended multiversion B-tree that offers the same asymptotic performance for every snapshot as an ordinary B-tree. In particular, the cMVBT has substantially redesigned the MVBT, introducing a new page layout, proactive splits, and most importantly, a novel multiversion concurrency control protocol that combines CAS and optimistic latching. As a unique result, range scans run latch-free and without having conflicts with write operations. This is especially beneficial when using optimistic latching, as it reduces the number of failed write operations that must be restarted. Scan operations never fail because they run only on immutable data. 

The results of an experimental evaluation demonstrate the consistent superiority of the cMVBT over B-trees with version chains for both scan and OLTP throughput across various workloads, where write operations comprise insertions, deletions, and updates. The lowest superiority over an advanced version-list approach is observed for updates only. For workloads including insertions and deletions, the differences can be orders of magnitude. 

Overall, we conclude that the cMVBT offers high concurrency on an efficient multiversion index structure, making it well-suited to HTAP workloads that require both high transactional throughput and efficient analytical scans.

%Thus, we conclude that the cMVBT is a highly suitable approach with excellent performance.  

% The results demonstrate that the dominant factor influencing query latency is the underlying data organization.

%The cMVBT achieves its performance advantages through three key design properties. First, data is organized by version in contiguous, cache-friendly blocks, avoiding pointer-heavy traversal and reducing cache misses. 
%Second, readers operate exclusively on immutable data, eliminating synchronization overhead and enabling efficient concurrent access. 
%Third, the cMVBT’s version-root organization naturally partitions the data into multiple small active subtrees, allowing OLTP transactions to operate on a compact live working set and thereby reducing traversal and contention costs.

%In addition, the cMVBT integrates a lightweight garbage collection mechanism that maintains stable memory consumption by reclaiming obsolete versions and retaining only relevant data. Reclaimed nodes are recycled internally, reducing allocation and de-allocation overhead and further improving cache locality and execution efficiency. Importantly, the GC introduces negligible runtime overhead even under mixed workloads.

% Hier brauchen wir nicht viel zu sagen
In our ongoing work, we extend the cMVBT to manage data stored on external storage and integrate it into a storage engine that supports multiversion transactions comprising multiple operations. Furthermore, we plan to examine compression techniques and a more advanced node design to improve traversals. 

% exploring disk-based nodes would allow the structure to scale beyond main memory and provide persistence, potentially integrating with traditional storage engines.

% examine how the cMVBT can serve as a concurrent storage structure for multiversion transactions that support a mix of operations. 

%to provide full transactional support under snapshot isolation, enabling multi-snapshot atomicity and consistent visibility across concurrent transactions. Finally, exploring disk-based nodes would allow the structure to scale beyond main memory and provide persistence, potentially integrating with traditional storage engines.

%Several directions remain for extending the current cMVBT implementation. First, incorporating page-level garbage collection would allow reclaiming space from obsolete versions more efficiently. Second, while our current design supports atomic operations within individual snapshots, a natural next step is to provide full transactional support under snapshot isolation, enabling multi-snapshot atomicity and consistent visibility across concurrent transactions. Finally, exploring disk-based nodes would allow the structure to scale beyond main memory and provide persistence, potentially integrating with traditional storage engines.
% \begin{acks}
%  This work was supported by the [...] Research Fund of [...] (Number [...]). Additional funding was provided by [...] and [...]. We also thank [...] for contributing [...].
% \end{acks}

%\clearpage

\bibliographystyle{ACM-Reference-Format}
\bibliography{sample}

\end{document}